%% file: log-survey.tex
\documentclass{amsproc}

\usepackage[latin1]{inputenc}
\usepackage[T1]{fontenc}
\usepackage{ae}
\usepackage{amsmath,amssymb}
\usepackage{eucal}
\usepackage{url}
\usepackage[cmtip,arrow]{xy}
\usepackage{pb-diagram,pb-xy}
\usepackage{graphicx}

\newtheorem{theorem}{Theorem}
\theoremstyle{definition}
\newtheorem{definition}[theorem]{Definition}
\newtheorem{algorithm}[theorem]{Algorithm}

\newcommand{\inoutput}[2]{\phantom{X} \quad \\
   \begin{tabular}{ll}\textsc{Input: } & #1\\
   \textsc{Output: } & #2 \end{tabular}}
\newcommand{\algif}{\textbf{if }}
\newcommand{\algrepeat}{\textbf{repeat }}
\newcommand{\alguntil}{\textbf{until }}

\newcommand{\zitatautor}[2]{\hfill{\it \begin{tabular}{l} #1
\end{tabular}} \par
\hfill --- \textsc{#2} \vspace{1cm}}
\newcommand{\F}{\mathbb{F}}
\newcommand{\M}{\mathbb{M}}
\newcommand{\N}{\mathbb{N}}
\newcommand{\R}{\mathbb{R}}
\newcommand{\Z}{\mathbb{Z}}
\newcommand{\Cc}{\mathcal{C}}
\newcommand{\Fc}{\mathcal{F}}
\newcommand{\Mcal}{\mathcal{M}}
\newcommand{\Pc}{\mathcal{P}}
\newcommand{\f}{\mathfrak{f}}
\newcommand{\Div}{\operatorname{Div}}
\newcommand{\Prin}{\operatorname{Prin}}
\newcommand{\J}{\operatorname{J}}
\newcommand{\Pic}{\operatorname{Pic}}
\newcommand{\rquo}[2]{#1 / #2}

\copyrightinfo{2007}{Andreas Enge}
\title{Discrete logarithms in curves over finite fields}
\author{Andreas Enge}
\address{INRIA Futurs \& Laboratoire d'Informatique (CNRS/UMR 7161),
\'Ecole polytechnique, 91128 Palaiseau Cedex, France}
\email{enge@lix.polytechnique.fr}
\date{December 23, 2007}

\begin{document}
\maketitle

\zitatautor {\textsc {Logarithme}, s. m. (Arithmét.)
nombre d'une progression \\ arithmétique,
lequel répond à un autre nombre dans une \\ progression géométrique.}
{Encyclopedia of Diderot and d'Alembert}

The discrete logarithm problem in finite groups is one of the supposedly difficult problems at the foundation of asymmetric or public key cryptography. The first cryptosystems based on discrete logarithms were implemented in the multiplicative groups of finite fields, in which the discrete logarithm problem turned out to be easier than one would wish, just as the factorisation problem at the heart of RSA. The focus has then shifted towards elliptic and more complex algebraic curves over finite fields. Elliptic curves have essentially resisted all cryptanalytic efforts and to date yield the cryptosystems relying on a number theoretic complexity assumption with the shortest key lengths for a given security level, while other classes of curves have turned out to be substantially weaker. This survey presents the history and state of the art of algorithms for computing discrete logarithms in non-elliptic curves over finite fields; the case of elliptic curves is touched upon, but a thorough treatment would require an article of its own, see \cite[Chapter~V]{bss99} and \cite{hes05}. For a previous survey on hyperelliptic curves in cryptography, including the discrete logarithm problem, see \cite{gau05hyp}.

Let us fix the notation used in the following. Given a cyclic group $(G, +)$ of order $N$, generated by some element $P$, the \textit {discrete logarithm} of $Q \in G$ to the base $P$ is given by the integer $x = \log Q = \log_P Q$, uniquely determined modulo $N$, such that $Q = xP$. The \textit {discrete logarithm problem (DLP)} in $G$ is to compute $x$ given $Q$. A cryptosystem is said to be \textit {based on the discrete logarithm problem in $G$} if computing discrete logarithms in $G$ breaks the cryptosystem (in some specified sense). Note that it is usually unknown whether breaking the system is indeed \textit {equivalent} to the discrete logarithm problem (but see the treatment of the computational Diffie--Hellman problem in Section~\ref {ssec:lower bounds}).

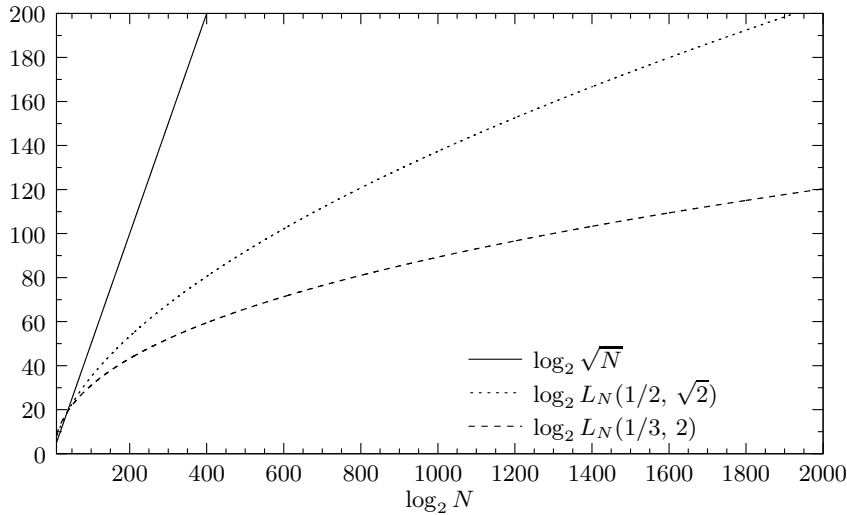
\begin {figure}[hbt]
\begin {center}
\input{function.pstex_t}
\end {center}
\caption {\label {fig}Complexity of the DLP in different groups}
\end {figure}

Figure~\ref {fig} illustrates the complexity of the discrete logarithm problem depending on $N$, as it presents itself in a number of groups suggested for cryptographic use. In the following sections, we will examine more closely algorithms in each of the complexity classes, going from the slower to the faster ones, that at the same time apply to a more and more restricted class of groups.

\section {Exponential algorithms}

\subsection {Generic algorithms}
\label {ssec:generic}

A certain number of algorithms allows to compute discrete logarithms generically using only group operations in $G$, independently of the concrete representation of its elements, under the only assumption that each group element is represented canonically by a unique bit string.

Note first that the decisional version of the DLP is easy: Given $Q$ and a candidate $x$ for its discrete logarithm, it suffices to compute $xP$ and to compare it to $Q$ in order to check whether $x$ is the correct logarithm. Hence the DLP may be solved with $O (N)$ group operations by exhaustive search.

This complexity may be reduced to $O (\sqrt N)$. Shanks's \textit {baby-step giant-step} algorithm \cite{sha72} computes first the baby steps $iP$ for $0 \leqslant i < \lceil \sqrt N \rceil$ and stores them in a hash table; then the giant steps $Q - j \lceil \sqrt N \rceil$ for $0 \leqslant j < \lceil \sqrt N \rceil$ are computed and looked up in the hash table containing the baby steps. As soon as a \textit {collision} $iP = Q - j \lceil \sqrt N \rceil P$ occurs, the discrete logarithm $x = i + j \lceil \sqrt N \rceil$ is readily deduced. This deterministic algorithm performs $O (\sqrt N)$ group operations and requires storage space for $O (\sqrt N)$ elements.

A probabilistic approach due to Pollard \cite{pol78} allows to dispense with the storage requirements. The basic idea is to compute random linear combinations $R_i = a_i P + b_i Q$. When a collision $R_i = R_j$ occurs, the discrete logarithm is given by $x = - \frac {a_j - a_i}{b_j - b_i} \bmod N$ if $b_j - b_i$ is invertible modulo $N$; otherwise, at least the partial information $x \bmod \frac {N}{\gcd (N, b_j - b_i)}$ is obtained. As such, the algorithm has an expected running time of $O (\sqrt N)$, but still needs to store $O (\sqrt N)$ group elements. By replacing the random choice of $a_i$ and $b_i$ by a pseudo-random walk such that $R_{i+1}$ depends only on $R_i$, and by looking for collisions exclusively of the form $R_i = R_{2i}$, one recovers Pollard's $\varrho$ algorithm, that heuristically takes a time of $O (\sqrt N)$ with constant storage. For a more advanced analysis, see \cite{tes01}.

Alternatively, one may use an approach admitting a simple parallelisation due to van Oorschot and Wiener \cite{ow99}. First of all, \textit {distinguished points} are defined as group elements with an easily recognisable property that occurs with a well-controlled probability, such as a certain number of zeroes in their binary representation. Several pseudo-random walks are started in parallel from different points. As soon as a distinguished point is reached, it is reported to a central machine that stores it and performs the collision search on only the stored elements.

The existence of a canonical representative for each element is crucial for the algorithms of complexity $O (\sqrt N)$; it allows to store the elements in a hash table and to perform a search in essentially constant time. If the collision search required a test for equality with each of the stored elements, the complexity would raise again to $O (N)$.

A classical trick described in \cite{ph78} consists in reducing the DLP to a series of discrete logarithm computations in the subgroups of order $p$ of $G$ for primes $p$ dividing $N$. First of all, if $e$ is the largest exponent such that $p^e | N$, one has $x \bmod p^e = \log_{(N / p^e) \cdot P} \left( \frac {N}{p^e} Q \right)$; the Chinese remainder theorem allows to compose these discrete logarithms in the Sylow subgroups of $G$ to obtain $x$. So without loss of generality, we may henceforth assume that $N = p^e$ is a prime power. Similarly, $x_0 = x \bmod p$ is obtained as $\log_{p^{e-1} P} (p^{e-1} Q)$; then, $x_1 = \frac {x - x_0}{p} \bmod p$ as $\log_{p^{e-1} P} (p^{e-2} (Q - x_0 P))$ and so on, so that the decomposition in base $p$ of $x$ is computed via a series of discrete logarithms in the subgroup of order $p$ of $G$. Combined with the algorithms of square root complexity presented above, discrete logarithms may be computed with
\[
O \left( \sum_{p^e || N} e \sqrt p \right)
\]
group operations, where the sum is taken over all prime powers $p^e$ such that $p^e | N$ and $p^{e+1} \nmid N$. So the maximal level of security reachable in a cryptosystem based on the discrete logarithm problem depends essentially on the largest prime factor of the group order. For prime $N$, this corresponds to the straight line in Figure~\ref {fig}.

\subsection {Lower bounds}
\label {ssec:lower bounds}

It would be interesting to know if a minimal difficulty of the DLP may be guaranteed. Nechaev and Shoup provide a partial answer in \cite{nec94,sho97}: If the only operations permitted are additions in the group and $N$ is prime, then $\Omega (\sqrt N)$ operations are needed to compute discrete logarithms with a non-negligible probability. To bypass this lower bound, an algorithm needs to take into account the particular representation of the group elements that distinguishes $G$ from the abstract cyclic group of order $N$.

Let us make a digression and briefly discuss the computational Diffie--Hellman problem (CDH), that is known to be equivalent to the security of a certain number of cryptosystems. It consists in computing $abP$ given the group elements $P$, $aP$ and $bP$. In the same article \cite{sho97}, Shoup shows that a generic algorithm for CDH requires $\Omega (\sqrt N)$ operations in a group of prime cardinality $N$. Even the decisional Diffie--Hellman problem DDH (given $P$, $aP$, $bQ$ and a candidate $Q$, decide whether $Q = abP$) has the same minimal complexity in this setting.

Maurer and Wolf have shown the equivalence between CDH and DLP independently of their difficulty in \cite{mw99rel}. Trivially, being able to solve DLP leads to solving CDH. For the converse direction, the authors consider the case that $N$ is a prime such that there is an auxiliary group $H$, algebraic over $\F_N$, into which $\F_N$ may be embedded in a probabilistic sense (the image of an element need not exist, but in this case, a slightly perturbed element must have an image). For instance, $H$ may be an elliptic curve defined over $\F_N$, and the image of $x \in \F_N$ is given by a point on $H$ with abscissa $x$, if it exists; otherwise, one may continue with $x + e$ for a small integer $e$. If in this situation the DLP in $H$ can be solved by an algebraic algorithm carrying out $n$ group operations, the DLP in $G$ can be solved by an algorithm making essentially $n$ calls to an oracle for CDH in $G$. If the order of $H$ is sufficiently smooth (that is, it does not have prime factors exceeding a polynomial in $\log N$), the algorithms of Section~\ref {ssec:generic} applied to $H$ lead to a polynomial time equivalence between CDH and DLP in $G$. Notice that only the order of $G$ plays a role in this argument, but not the concrete representation of its elements. Using the fact that all integers in the Hasse interval around $N$ appear as cardinalities of elliptic curves over $\F_N$ and assuming heuristically that they have the same factorisation pattern as random integers of the same size, Maurer and Wolf show the existence of an auxiliary group $H$ such that the reduction becomes polynomial. Finding the group via complex multiplication, however, may take exponential time. If a subexponential reduction in $L (1/2)$ is considered sufficient instead of a polynomial reduction, it should be possible to find the group in the same subexponential time.

The generic, exponential algorithms are for the time being the only ones that may be applied to arbitrary elliptic curves. Some particular elliptic curves admit an embedding into another group in which discrete logarithms are considerably easier to compute, but these curves have a very low density: They are supersingular and other curves with a low embedding degree into the multiplicative group of a finite field \cite{mov93,fr94}; subgroups of order $p$ defined over a finite field $\F_q$ of characteristic $p$, that may be embedded into the additive group $(\F_q, +)$ \cite{sa98,sem98eva,sma99}; and elliptic curves that may be embedded by Weil descent into the Jacobian of a hyperelliptic curve of low genus following an idea suggested by Frey in \cite{fre01}, see \cite{ghs02con,die03} and \cite{hes05} and the references therein. Weil descent provides another motivation for examining more closely curves of genus larger than~$1$.

\section {Subexponential algorithms of complexity $L (1/2)$}

\subsection {The subexponential function}

Under the designation \textit {subexponential function}, one might subsume all functions that grow more slowly than exponentially, but faster than polynomially. In the context of discrete logarithm or factorisation algorithms, the following more restrictive definition appears naturally:

\begin {definition}
\label {def:subexp}
The {\em subexponential function} with parameters $\alpha \in (0, 1)$ and $c > 0$ with respect to the argument $N$ is given by
\[
L_N (\alpha, c) = e^{c (\log N)^\alpha (\log \log N)^{1 - \alpha}}.
\]
To simplify the notation, we let
\[
L_N (\alpha) = \left\{ L_N (\alpha, c) : c > 0 \right\}
\]
and omit the subscript $N$ when it is understood from the context.
\end {definition}

In the following, we will focus on the parameter $\alpha$, that has the biggest influence on the growth of the function. The parameter $c$ is often called the \textit {constant} of the subexponential function, although it appears in fact in the exponent, so that its influence is far from negligible.

The traditional notation $L_N$ may lead to confusion, as in terms of complexity, one has to assume that a problem with $N$ possible inputs is specified by $\log N$ bits, and  subexponentiality has to be understood with respect to $\log N$ rather than $N$:
\begin {itemize}
\item
The extreme case $\alpha = 0$, excluded by the definition, leads to the polynomial
$\log^c N$;
\item
the other extreme case $\alpha = 1$ leads to the exponential $N^c$;
\item
as intermediate values, essentially $\alpha = 1/2$ and $\alpha = 1/3$ occur in the contexts of discrete logarithms and of factorisation. Two typical functions are traced in Figure~\ref {fig}.
\end {itemize}
The following computation rules are easily checked:
 \begin {equation}
\label {eq:subexp}
\begin {split}
& L_N (\alpha, c_1) \cdot L_N (\alpha, c_2) = L_N (\alpha, c_1 + c_2) \\
& \log^k N \in L (\alpha, o (1)) \text { for any $k$, and more generally,}
\\
& L_N (\beta, d) \in L_N (\alpha, o (1)) \text { for } \beta < \alpha.
\end {split}
\end {equation}
In particular, if a polynomial time operation is repeated $L_N (\alpha, c)$ times, the resulting complexity is in $L_N (\alpha, c + o (1))$; this is why Definition~\ref {def:subexp} is often modified to allow a $o (1)$ term in the constant.

\subsection {An algorithm for finite fields}

Subexponential algorithms for discrete logarithms usually proceed in two stages: In the first stage, called \textit {sieving} or \textit {relation collection}, an integral matrix is filled with \textit {relations}; the \textit {linear algebra} stage solves the resulting system modulo the group order and yields the discrete logarithms of certain elements; a third, comparatively inexpensive stage may be needed to compute \textit {individual logarithms}. It has become common to call this kind of algorithm ``index calculus'', a rather unfortunate terminology, since ``index'' is traditionally used as a synonym for ``logarithm''. Already the encyclopedia by Diderot and d'Alembert, published between 1751 and 1772, gives the following definition: ``\textit {Index}, en terme d'Arithm\'etique, est la m\^eme chose que la caract\'eristique ou l'exposant d'un logarithme. Voyez \textsc {Logarithme}.'' \cite{da51}.

The basic idea of creating relations and of combining them linearly for computing discrete logarithms (and for factoring) has been published by Kra{\"\i}tchik in the twenties \cite[Chapter~5, \S\S14-16]{kra22}. In 1979, the algorithm has been rediscovered by Adleman and presented with the analysis of its subexponential complexity for the case of finite prime fields. It is easily generalised to $\F_{2^m}$ (for the reasons explained in Section~\ref {ssec:framework}). In the following, we describe a slightly modified version.

To recall the problem, let $P$ be a primitive element of $\F_{2^m}$ and $Q \in \F_{2^m}^\times$; we wish to compute $x$ such that $Q = P^x$. The finite field is conveniently represented as $\F_2 [X] / (f)$ with $f$ an irreducible polynomial of degree $m$, such that an element of $\F_{2^m}$ may be considered as a binary polynomial of degree less than $m$. This representation of the field elements by polynomials introduces notions that in principle have no meaning in a field: It is now possible to speak of \textit {irreducible elements}, the degree of the polynomials leads to a notion of \textit {size} of the elements, and there is a \textit {unique factorisation} of elements into irreducibles. In fact, the factorisation is no more unique as soon as the restriction on the degrees is lifted, as several elements of $\F_2 [X]$ may represent the same element of the finite field.

\begin {algorithm}
\label {alg:l12}
\inoutput {$P$ a primitive element of $\F_{2^m} = \F_2 [X] / (f)$, $Q \in \F_{2^m}^\times$}{$x$ such that $Q = P^x$}
\begin {enumerate}
\setcounter{enumi}{-1}
\item
Let $N = 2^m-1$. Fix a {\em smoothness bound} $B \in \N$ and compute the {\em factor base} $\Fc = \{ p_0, \ldots, p_n \}$ containing the irreducible polynomials over $\F_2$ of degree at most $B$ and $P = p_0$. Prepare an empty matrix $A$ with $n$ columns and $r$ rows and an empty vector $b$ with $r$ rows for some $r$ slightly larger than $n$.
\item
\algrepeat for $i = 1, \ldots, r$ \\
\hspace*  {5mm} \algrepeat \\
\hspace* {10mm} draw random exponents $e_{i j} \in \{ 0, \ldots, N-1 \}$
               for $j = 0, \ldots, n$ \\
\hspace* {10mm} compute $\prod_{j=0}^n p_j^{e_{ij}} \bmod f$ \\
\hspace* {10mm} \algif the result factors over $\Fc$ as $\prod_{j=0}^n p_j^{f_{ij}}$ \\
\hspace* {15mm} there is a {\em relation} $\prod_{j=1}^n p_j^{a_{ij}} = P^{-a_{i0}}$
               in $\F_{2^m}$ with $a_{ij} = e_{ij} - f_{ij}$ \\
\hspace* {15mm} add $(a_{ij})_{j=1}^n$ to the matrix $A$ and $-a_{i0}$ to the
               vector $b$\\
\hspace*  {5mm} \alguntil success in creating a new relation
\item
Solve the linear system $A y = b$ modulo $N$, so that $y_j = \log_P p_j$.
\item
Create an additional relation $Q \prod p_j^{e_j} = \prod p_j^{f_j}$ as above;
return $x = \sum (f_j - e_j) y_j$. 
\end {enumerate}
\end {algorithm}

This version of the algorithm separates the linear algebra of stage~2 and the computation of individual logarithms of stage~3, that may be repeated as many times as desired. Alternatively, it would be enough to add the target $Q$ to the factor base and to stop after stage~2.

The complexity of the algorithm depends essentially on the probability that a polynomial of degree at most $m-1$ decomposes completely over the factor base, otherwise said, that it is \textit {$B$-smooth}. If the factor base size $n$ is polynomial, this probability decreases exponentially; for it to decrease polynomially, one would need $n$ to be exponential. The optimal value is somewhere in between; precisely, for $B$ chosen such that $n \in L (1/2)$, the probability of obtaining a relation is in $1 / L (1/2)$ (see \cite{bp98} and Section~\ref {ssec:framework}). Hence, the expected number of iterations for obtaining one relation is in $L (1/2)$, and the process has to be repeated $r \in O (n) \subseteq L (1/2)$ times to fill the matrix. As all the basic operations of the algorithm are polynomial or in $L (1/2)$ (for instance, the linear algebra stage is polynomial in $n$), the computation rules (\ref {eq:subexp}) show that the total complexity of the algorithm is in $L (1/2)$. Smoothness is discussed in more generality and a more detailed complexity analysis is developed in Section~\ref{ssec:framework}.

Notice that the subexponential complexity of Algorithm~\ref {alg:l12} does not contradict the exponential lower bounds of Section~\ref {ssec:lower bounds}: We clearly make use of the particular representation of the elements of $\F_{2^m}^\times \simeq \Z / N \Z$ by polynomials, and the algorithm is far from generic.

\subsection {Arithmetic of Jacobians}
\label {ssec:jacobians}

In the light of Section~\ref {ssec:lower bounds}, it is clear that we need to take a closer look at the representation of elements and at the group law associated to an algebraic curve. To arrive at the analogue of Algorithm~\ref {alg:l12}, our aim is to show that these elements behave essentially like polynomials.

\subsubsection* {Curves over algebraically closed fields}

For the time being, let us consider a curve $\Cc$ defined by a non-singular irreducible polynomial $C (X, Y)$ over an \textit {algebraically closed} field $K$. The \textit {points} on $\Cc$ are the $(x, y) \in K^2$ such that $C (x, y) = 0$. Non-singularity means that for no point $(x,y)$ on the curve, the partial derivatives $\frac {\partial C}{\partial X} (x, y)$ and $\frac {\partial C}{\partial Y} (x, y)$ vanish simultaneously. There is an integer $g$, called the \textit {genus} of the curve, that is closely related to the degree of $C$ if the equation is ``reasonable'', and that measures, roughly speaking, how ``complicated'' the curve is. For instance, a \textit {hyperelliptic} curve in characteristic different from~$2$ is given by a non-singular polynomial $C = Y^2 - X^{2g+1} - f (X)$ with $f$ of degree at most $2g$; the case $g=1$ is that of an \textit {elliptic} curve. A more general case is that of \textit {superelliptic} curves, defined by a non-singular polynomial $Y^a - X^b - f(X)$ with $f$ of degree less than $b$ and $\gcd (a, b) = 1$ when the characteristic of $K$ is coprime with $a$. By admitting certain mixed terms, one obtains the most general curves that have been suggested for use in cryptography, namely $\Cc_{a,b}$ curves, given by a non-singular irreducible polynomial of the form
\[
Y^a - X^b - \sum_{(i, j) : a i + b j < a b} c_{ij} X^i Y^j
\]
with $\gcd (a, b) = 1$ when the characteristic of $K$ divides neither $a$, nor $b$. The genus of these curves is given by $g = \frac {(a-1)(b-1)}{2}$.

To a curve, one can associate its \textit {coordinate ring} $K [C] = K [X, Y] / (C)$, the ring of polynomial functions from the curve to the field $K$. In the case of a $\Cc_{a, b}$ curve, $K [\Cc]$ can be seen as the set of polynomials of arbitrary degree in $X$ and of degree at most $a-1$ in $Y$, since each occurrence of $Y^a$ may be replaced by $X^b + \sum c_{ij} X^i Y^j$. The field of fractions of $K [\Cc]$ is denoted by $K (\Cc)$ and is called the \textit {function field} of~$\Cc$; it consists of the rational functions from the curve to $K$. 

Except for elliptic curves, the associated group does not consist of only the points on the curve. Instead, one has to consider the \textit {Jacobian} $\J (\Cc)$ of the curve, an abelian variety. In practice, it is preferable to work with the isomorphic group (denoted by $\Pic^0 (\Cc)$ or again by $\J (\Cc)$) of divisor classes of degree~$0$. Define the group of \textit {divisors} of $\Cc$ by
\[
\Div (\Cc) = \left\{ \sum_{P \in \Cc} m_P P : m_P \in \Z, \text { almost all zero} \right\},
\]
the set of finite formal sums of points with potentially negative coefficients. This definition is in fact slightly wrong; instead of only considering points on the affine curve, one needs to also take into account ``points at infinity'' on the projective closure of $\Cc$. Moreover, the projective closure will usually be singular at infinity; instead of a singular point, one needs to consider several points corresponding to its resolution on a non-singular model. Equivalently, one may define divisors as formal sums of places of the function field $K (\Cc)$ instead of points. Function fields of hyperelliptic or, more generally, $C_{a, b}$ curves are particular in that they have only one place at infinity; so it suffices to augment the set of points by one additional special point called $\infty$.

The \textit {degree} of a divisor is given by
\[
\deg \left( \sum m_P P \right) = \sum m_P.
\]
The degree~$1$ divisors containing only one point with multiplicity~$1$ are called \textit {prime}; indeed, they constitute indivisible atoms, and any divisor may be written uniquely as a sum of prime divisors.

Associate to a rational function $f \in K (\Cc)$ its \textit {principal divisor}, containing its zeroes with positive and its poles with negative multiplicities. At $\infty$ on a $C_{a,b}$ curve, the function $X$ has a pole of order $b$, the function $Y$ a pole of order $a$; the order at $\infty$ of any other function may be deduced by the triangular inequality. It turns out that the degree of a principal divisor is~$0$, otherwise said, a rational function has as many zeroes as poles, counting multiplicities. Let $\Div^0 (\Cc)$ denote the group of degree~$0$ divisors and $\Prin (\Cc)$ its subgroup of principal divisors; then the Jacobian is given by
\[
\J (\Cc) = \Div^0 (\Cc) / \Prin (\Cc).
\]
Given $\infty$ (or any other point on the curve, for that matter), there is a natural isomorphism
\[
\Div^0 (\Cc) \to \Div' (\Cc), \quad
\sum_P m_P P \mapsto \sum_{P \neq \infty} m_P P,
\]
where $\Div' (\Cc)$ denotes the subgroup of $\Div (\Cc)$ of divisors not containing $\infty$ in their support. The inverse isomorphism is obtained by adding the right multiple of $\infty$ to obtain a degree~$0$ divisor. By the Riemann--Roch theorem, each class of $\J (\Cc)$ can then be represented by a unique \textit {effective} or \textit {positive} divisor (that is, without negative coefficients) in $\Div' (\Cc)$ of minimal degree, which is called its \textit {reduced (along~$\infty$) representative}. Its degree is moreover bounded by the genus~$g$.

The coordinate ring $K [\Cc]$ is in fact the set of functions without poles at infinity (or otherwise said, the integral closure of $K [X]$ in $K (\Cc)$). But since for $\Cc_{a, b}$ curves, $\infty$ is the only point at infinity, this implies that the affine points on the curve are in bijection with the prime ideals of $K [\Cc]$, that $\Div' (\Cc)$ is isomorphic to the group of fractional ideals of $K [\Cc]$ and that $\J (\Cc)$ is isomorphic to the ideal class group of $K [\Cc]$. This observation allows to switch to the standard representation of ideals in extensions of Dedekind domains: Any divisor $D \in \Div' (\Cc)$ may be represented by an ideal of $K [\Cc]$ in the form
\begin {equation}
\label {eq:divisor}
D = (d)(u, w)
\end {equation}
with $d$, $u \in K [X]$ monic and $w \in K [\Cc]$ (cf. \cite[\S~163, p.~461]{dd71} or \cite[Th.~17]{mar77} for a proof in the number field case). The polynomial $w$ may be taken to be monic and, for a $\Cc_{a,b}$ curve, of degree less than $a$ in $Y$. Since $(d)$ is principal, any element of the Jacobian is represented as $(u, w)$. Even without recourse to the theory of Dedekind rings, the existence of such a representative may be shown by choosing $u \in K [X]$ having as zeroes (with the right multiplicities) the $X$-coordinates of the points in the divisor $D$, and by letting the bivariate polynomial $w$ interpolate (again with the correct multiplicities, which requires some care) the $Y$-coordinates. Notice that a prime divisor $P = (x, y)$ is characterised by a representative, namely $(X - x, Y - y)$, in which the first polynomial is irreducible.

Relying on the representation (\ref {eq:divisor}) of divisors, the algorithm realising the group law in a Jacobian works with polynomials and proceeds in two steps: The \textit {composition} step corresponds to the addition of the divisors respectively the multiplication of the ideals, while taking out principal ideals $(d)$ of $K [X]$ that may appear; essentially, this is Lagrangian interpolation. The \textit {reduction} step computes for the resulting divisor, that is generically of degree $2g$, its unique reduced representative of degree at most $g$; this part of the algorithm depends heavily on the curve. Efficient algorithms for arbitrary curves have been developed by He{\ss} and Khuri-Makdisi \cite{hes02com,khu04}. For hyperelliptic curves, see, for instance, \cite{can87,eng01ext,lan05,gau07}; for superelliptic curves, see \cite{gps02,befg05}, for $\Cc_{a,b}$ and in particular $\Cc_{3,4}$ curves, see \cite{ari99alg,ari03,fo04,befg04,ak07}.

\subsubsection* {Curves over finite fields}

In order to obtain finite groups, it is clearly necessary that the curves under consideration be defined over a finite field $K = \F_q$. Contrarily to what one might expect, it is not sufficient to emulate the construction made for algebraically closed fields: When adding two elements containing only points defined over $\F_q$, the reduction may result in a divisor containing points defined over an extension field. To save the situation, one needs to resort to Galois invariance and consider the Frobenius automorphism of $\F_q$, $x \mapsto x^q$, that yields naturally the endomorphism $\varphi : (x, y) \mapsto (x^q, y^q)$ on the points of the curve defined over $\overline \F_q$. It is trivially extended to divisors and rational functions. The groups $\Div$, $\Div^0$ and $\Prin$ may thus be defined as above as sets of divisors defined over $\overline \F_q$, but with the additional restriction that they be invariant under $\varphi$. To obtain $\Div'$ in an analogous manner, one furthermore needs $\infty$ to be defined over $\F_q$, which is the case for all curves under consideration. So once again, we end up with the ideal class group of $K [\Cc]$. The elements of the Jacobian are represented as above by ideals $(u, w)$, now with $u$ and $w$ having coefficients in $\F_q$. The algorithms of composition and reduction remain unchanged; their algebraic nature implies that they have no ``conscience'' of the field over which they work.

By Weil's theorem \cite{wei71}, the order of the Jacobian of a curve $\Cc$ of genus $g$ defined over $\F_q$ satisfies
\begin {equation}
\label {eq:weil}
(\sqrt q - 1)^{2g} \leqslant |\J (\Cc)| \leqslant (\sqrt q + 1)^{2g}.
\end {equation}

\subsubsection* {Composition and decomposition}

The previous discussion shows that the arithmetic of the Jacobian groups of the curves under consideration boils down to that of bivariate polynomials. But as far as discrete logarithms are concerned, the group elements even behave essentially like univariate polynomials.

As a consequence of Weil's theorem, the majority of elements of the Jacobian is represented by $(u, w)$ with $\deg u = g$, $w = Y - v (X)$ and $\deg v = g - 1$. When two distinct elements $D_1 = (u_1, Y - v_1)$ and $D_2 = (u_2, Y - v_2)$ are to be added, with overwhelming probability one has $\gcd (u_1, u_2) = 1$, or otherwise said, the points in $D_1$ have distinct $X$-coordinates from those in $D_2$. Then the result of the composition is $D_1 + D_2 = (u, Y - v)$ with $u = u_1 u_2$ and $v$ the Lagrangian interpolation polynomial such that $v_i = v \bmod u_i$, which is in fact independent of the curve. The composition step for doubling a divisor $(u_1, Y - v_1)$ usually results in $(u, Y - v)$ with $u = u_1^2$ and $v$ a Hensel lift (that depends on the curve). As long as $\deg u$ does not exceed $g$, in general no reduction occurs. Otherwise, the reduction step is also specific to the curve. So adding divisors corresponds to multiplying the $u$-polynomials in $\F_q [X]$ and updating the $v$-polynomials accordingly, followed by a reduction step. In this sense, the addition in the Jacobian behaves like multiplication in $\F_q^{g+1}$, which also proceeds by multiplying polynomials of degree at most~$g$, followed by a reduction.

Over a finite field $\F_q$, a \textit {prime divisor} is a divisor that cannot be written as a sum of two non-trivial divisors defined over the same field. Concretely, a prime divisor of degree $k$ is given by the orbit $D$ under the Galois endomorphism of a point $P = (x, y)$ with coordinates $x$ and $y$ in $\F_{q^k}$, but not both in the same subfield. A typical representative occurs when $x$ itself is not defined in a subfield of $\F_{q^k}$; then $D = (u, Y - v)$ with $u$ the minimal polynomial of degree $k$ of $x$. (The other prime divisors have the form $(u, w)$ with $\deg u$ a proper divisor of $k$ and $\deg_Y w = k / \deg u$ and occur in negligible numbers.) In any case, prime divisors are again characterised by having an irreducible first polynomial.

As the decomposition of an element of $\F_{q^{g+1}}$ in Algorithm~\ref {alg:l12}, the prime decomposition of a divisor of the form $D = (u, Y - v)$ also boils down to factoring a polynomial in $\F_q [X]$; if $u = \prod u_i^{e_i}$, then $D = \sum e_i D_i$ with $D_i = (u_i, Y - v \bmod u_i)$.

\subsection {Algorithms for hyperelliptic curves}
\label {ssec:l12hyperelliptic}

The first subexponential algorithm for computing discrete logarithms in hyperelliptic curves of large genus defined over a finite field $K = \F_q$ is due to Adleman, DeMarrais and Huang \cite{adh94}. It differs from Algorithm~\ref {alg:l12} essentially by the way in which the relations are created. Let the factor base $\Fc$ be given by all prime divisors of degree bounded by some smoothness bound $B$. Since principal divisors are zero in the Jacobian, it is sufficient to draw random polynomials of the form $Y - v (X)$ and to compute their divisors (higher degrees in $Y$ do not occur in hyperelliptic curves, since the polynomials may be reduced modulo the curve equation of degree~$2$ in $Y$). A smooth divisor, that is a divisor decomposing over $\Fc$, directly yields a relation. This is the case if the norm of $Y - v$ with respect to the function field extension $K (\Cc) / K (X)$, a polynomial in~$X$, is $B$-smooth. Assuming heuristically that norms behave like random polynomials of the same degree, the authors prove a complexity of $L_{q^g} (1/2)$ whenever $(2 g + 1)^{0.98} \geqslant \log q$. The result is heuristic for a second reason. Implicitly, Algorithm~\ref {alg:l12} describes an isomorphism between the group and $\Z^n$ modulo the lattice formed by the rows of the relation matrix. It is unclear whether the bounds one needs to impose on the degree of $v$ for a subexponential running time allow to obtain a sufficiently dense lattice to yield the isomorphism.

The first algorithm for discrete logarithms in hyperelliptic curves with a proven subexponential running time is given in \cite{eng02}. Essentially, it is Algorithm~\ref {alg:l12}, that applies directly to curves via the discussion at the end of Section~\ref {ssec:jacobians}. It relies on the fact, proved in \cite{es02}, that the proportion of smooth divisors is the same as the proportion of smooth univariate polynomials. (A similar result for the discrete logarithm problem in the infrastructure of a real quadratic function field can already be found in \cite{mst99}; it also relies on the smoothness theorem of \cite{es02}.) The constant of the subexponential complexity depends on the growth of the genus~$g$ with respect to the finite field size~$q$; precisely, a running time of
\[
L_{q^g} \left( 1/2,
\frac {5}{\sqrt 6} \left( \sqrt {1 + \frac {3}{2 \vartheta}} + \sqrt {\frac {3}{2 \vartheta}} \right)
+ o (1) \right)
\]
is proved in \cite{eng02} under the assumption that $g \geqslant \vartheta \log q$.

\subsection {A general framework}
\label {ssec:framework}

The similarities between finite fields, Jacobians of curves and other groups in which subexponential algorithms in $L (1/2)$ exist to solve the discrete logarithm problem, have motivated us to develop a framework that allows an abstract presentation and unified analysis independently of the group \cite{eg02}. It is explained in the following to give a more detailed complexity analysis for Jacobians of curves, as it is not more involved than a treatment of only the curve case.

Let $\Pc$ be a set of elements called \textit {prime}, and let $\M$ be the free monoid over $\Pc$. Suppose there is an equivalence relation $\sim$ in $\M$, compatible with addition, such that $G = \M / \!\!\sim$ is a group. Suppose furthermore that there is a \textit {size function} $\deg : \Pc \to \R^{\geqslant 1}$ (extended homomorphically to $\M$), which allows to define the factor base $\Fc$ as the set of prime elements of size bounded by a smoothness bound $B$. If the elements of $G$ have canonical representatives in $\M$ (usually, the smallest ones), the unique decomposability of elements of $\M$ into primes is inherited by $G$. If some technical conditions on $G$ concerning, for instance, the computability of the group law, the bit size of elements or the generation of $G$ by $\Fc$, are satisfied, then Algorithm~\ref {alg:l12} may be applied without modification to $G$.

These notions have been introduced by Knopfmacher in \cite{kno75}; he calls $\M$ an \textit {arithmetical semigroup} and $G$ an \textit {arithmetical formation}. Concrete examples are provided by prime fields $\F_p$, for which $\M = \Z$, $\deg$ is the logarithm and $\sim$ is equivalence modulo $p$; and finite fields $\F_{p^m} = \F_p [X] / (f)$, for which $\M = \F_p [X]$, $\deg$ is indeed the degree and $\sim$ is equivalence modulo $f$. But also class groups of number fields, for which $\M$ is the set of integral ideals, $\deg$ is the logarithm of the norm and $\sim$ is equivalence modulo prime ideals. And finally Jacobians of curves $\Cc$ over a finite field $\F_q$ with a unique point at infinity, for which $\M = \Div' (\Cc)$, $\deg$ is the degree of a divisor and $\sim$ is equivalence modulo principal divisors.

It remains to prove the running time of the algorithm. For it to be in $L (1/2)$, we need that a factor base of size $L (1/2)$ implies a smoothness probability of $1 / L (1/2)$. Corresponding results can be found, for instance, in \cite{pom87} for $\F_p$, in \cite{bp98} for $\F_{2^m}$, in \cite{sey87} for class groups of imaginary-quadratic number fields (under the generalised Riemann hypothesis) and in \cite{es02} for hyperelliptic curves of large genus.
Having a closer look at these smoothness theorems, one realises that they are essentially all the same: For a factor base of size $L_N (1/2, c)$, an element of size $\log N$ is smooth with a probability of $1 / L_N (1/2, 1 / (2c) + o (1))$. This result may be proved in $\M$ under an assumption analogous to the prime number theorem: The number of primes of size bounded by $k$ must be of the order of $\frac {q^k}{k}$ for some $q$, see \cite{man92sem,man92rem}. Then the number of elements of size at most $x$ that are smooth with respect to a bound $y$ is asymptotically (with some constraints on the respective growth of $x$ and $y$) given by the value of the Dickmann--de Bruijn function $\varrho$ in $u = \frac {x}{y}$; and de Bruijn has shown in \cite[(1.8)]{bru51asy} that
$1 / \varrho (u) \in e^{(1 + o (1)) u \log u}$, which provides the link with the subexponential function.

Due to the equivalence relation, the smoothness result for $\M$ cannot be directly transferred to $G$. In a curve, for instance, there are non-reduced divisors of degree~$g$, that as such do not occur as representatives of Jacobian elements. Nevertheless, the results of \cite{sey87,es02} provide examples of arithmetical formations in which the same smoothness behaviour may be observed; it is thus reasonable to accept it heuristically also in other contexts.

Given the smoothness result, the complexity of Algorithm~\ref {alg:l12} may be easily verified. Let $n = L_N (1/2, d)$ denote the size of the factor base, with $N$ the group order and $d$ a parameter to be determined later; in the curve case, $N$ must be replaced by $q^g$, which makes sense in the light of Weil's theorem (\ref {eq:weil}). If a group element may be decomposed over the factor base in time $L_N (1/2, o (1))$ (which is the case for all groups under consideration), the time to create $r = O (n)$ relations in the first stage of the algorithm is in
\[
L_N (1/2, o (1)) \cdot L_N (1/2, 1 / (2d) + o (1)) \cdot L_N (1/2, d)
= L_N (1/2, d + 1 / (2d) + o (1))
\]
by (\ref{eq:subexp}). The linear algebra stage treats a sparse matrix of order $L_N (1/2, d)$ with $L_N (1/2, d + o (1))$ entries, as each relation has on the order of $\log N$ coefficients. The Lanczos and Wiedemann algorithms of \cite{lan52,wie86} run in time $L_N (1/2, 2 d + o (1))$ on these matrices. Hence, the total running time of the algorithm becomes
\[
L_N (1/2, \max (d + 1 / (2d), d) + o (1)).
\]
This quantity is minimised by $d = \sqrt 2 / 2$, resulting in a complexity of
\[
L_N (1/2, \sqrt 2 + o (1)).
\]
For hyperelliptic curves, this running time holds when the field size $q$ remains fixed and the genus $g$ tends to infinity. When $q$ grows as well, the discrete nature of the degree function starts to play a role, but a moderate growth of $q$ may be tolerated (see also Section~\ref {ssec:l12low}); following the analysis of \cite{eng02}, a running time of
\[
L_N \left( 1/2, \sqrt 2 + \frac {2}{\vartheta} + o (1) \right)
\]
is obtained in \cite{eg02} for the case that $g \geqslant \vartheta \log q$.

An assumption that is implicit in Algorithm~\ref {alg:l12} need not be satisfied for curves: The group must by cyclic and of known order $N$. If this is not the case, one may replace the solution of a linear system by the computation of the Hermite and Smith normal forms of the matrix, which yields a complexity of $L_N (1/2)$ with a worse constant as, for instance, at the end of Section~\ref {ssec:l12hyperelliptic}; see \cite{eng01gen}.

An algorithm of proved subexponential complexity in $L_{q^g} (1/2 + \varepsilon)$ is given by Couveignes in \cite{cou01} for a large class of curves, not limited to hyperelliptic ones, under the mild assumption that the curve contains an $\F_q$-rational point and that the order of its Jacobian is bounded by $q^{g + O (\sqrt g)}$. The approach is quite different from the one presented here and relies on a double randomisation, of the combination of factor base elements as well as of the choice of a function in a certain Riemann--Roch space. An algorithm without restriction on the input curve is given by He{\ss} in \cite{hes04}, who thus proves a complexity of $L_{q^g} (1/2)$ for all curves of large genus.

\subsection {The low genus case}
\label {ssec:l12low}

At first sight, these algorithms do not seem to work in low genus. This is nicely illustrated by the case of elliptic curves, which are of genus~$1$, so that each reduced divisor contains exactly one point: Either, the smoothness bound is set to $B=1$, in which case any divisor is smooth, and the matrix contains as many columns as there are elements in the group, that is around $q$. So the algorithm must be slower than the generic ones of Section~\ref {ssec:generic}, of complexity $O (\sqrt q)$. Or the smoothness bound is set to $B=0$, in which case no divisor is smooth. The problem stems again from the discreteness of $B$, that is smoothed out when the genus becomes larger.

In fact, interesting results are already obtained for rather small genus, as first observed by Gaudry in \cite{gau00alg}. Assume $g$ to be fixed, while $q$ tends to infinity. Choose a smoothness bound of $B=1$, so that the factor base is composed of divisors containing only one point. By the Weil bound, its size $n$ satisfies $|n - (q + 1)| \leqslant 2 g \sqrt q = O (\sqrt q)$. The smooth reduced divisors are essentially the multisets containing $g$ points (cf. Section~\ref {ssec:jacobians}); asymptotically for $q \to \infty$, multiplicities do not play a role, so that the number of smooth reduced divisors is well approximated by $\binom {n}{g} = \frac {q^g}{g!} + O (q^{g - 1/2})$. As by (\ref {eq:weil}) the Jacobian group has a size of $q^g + O (q^{g - 1/2})$, the smoothness probability for a random element is, asymptotically for $q \to \infty$, given by $\frac {1}{g!}$. So filling the matrix requires $g! n = O (q)$ trials, and the linear algebra step on a sparse matrix with $O (n)$ rows and columns and $g$ entries per row takes $O (n^2 g) = O (q^2)$ arithmetic operations. While this complexity is exponential in the group size $q^g$, the generic algorithms of complexity $q^{g/2}$ are beaten as soon as $g \geqslant 5$. Notice that the two stages of the algorithm are quite unbalanced; by reducing the factor base size, an idea attributed to Harley in \cite{gau00alg}, one may slow down the relation collection process while speeding up the linear algebra. Letting $n = O (q^r)$ with $r = 1 - 1 / (g+1)$, the total complexity becomes $O (q^{2 - 2 / (g+1)})$ arithmetic operations, which is better than the generic complexity already for $g = 4$.

A further improvement is given by Thériault in \cite{the03}. Again, the factor base comprises only a fraction of the prime divisors of degree~$1$, chosen arbitrarily, say $n = q^r$ with some parameter $r$ to be optimised. The other prime divisors of degree~$1$, however, are not discarded any more, but form the set of \textit {large primes} (that in this context, of course, are not larger than the others, but the terminology as well as the basic idea is inspired by the large prime variation in the factorisation context). A relation is retained if it either consists of prime divisors in the factor base (the case of a \textit {full relation}) or if it contains exactly one large prime besides elements of the factor base (the case of a \textit {partial relation}). Before entering the linear algebra step, $k$ partial relations with the same large prime are combined to form $k-1$ full relations; large primes that occur only once are eliminated. The net effect is similar to the choice of a smaller factor base: Relation collection is slowed down (but not as much), while the linear algebra is accelerated. Thériault shows that the optimal value for $r$ is $1 - 2 / (2g + 1)$ for a final running time of $O (q^{2 - 4 / (2 g+1)})$ arithmetic operations; this is slightly better than the generic algorithms already for $g = 3$.

Finally, Gaudry, Thomé, Thériault and Diem in \cite{gttd07} and Nagao in \cite {nag07} have suggested the use of two ``large'' primes, which complicates the process of recombining partial relations, but allows to reduce the factor base size even further. The optimal value $r = 1 - 1 / g$ yields a running time of $O (q^{2 - 2/g})$ arithmetic operations.

The above algorithms are formulated for Jacobians of hyperelliptic curves, but carry over to arbitrary curves. One conclusion to draw might be that curves of genus~$3$ and above should be banned from cryptography, as they are less secure than lower genus curves for the same group order. As a more nuanced reaction, one may also decide to increase the group size slightly, especially for a genus close to the cross-over point. In genus~$3$, for instance, one would need to increase the bit length of the group order by $12.5~\%$ for an equivalent level of security compared to elliptic or genus~$2$ curves. This need not be penalising since machine word sizes introduce an effect of discretisation into the implementation.

\section {Subexponential algorithms of complexity $L (1/3)$}

Following the progress for factorisation algorithms, a complexity of $L (1/3)$ has also been established for discrete logarithm computations in finite fields. First of all, Coppersmith's algorithm \cite{cop84} treats $\F_{2^m}$; it may be seen as a special case of Adleman's \textit {function field sieve} \cite{adl94}, that applies to fields $\F_{p^m}$ with $p$ small. The case of $\F_p$ respectively $\F_{p^m}$ for $m$ small is handled by Gordon's \textit {number field sieve} \cite{gor93dis}. Recently, it has been shown in \cite{jlsv06} that the applicability domains of the two algorithms intersect, so that a complexity of $L (1/3)$ is obtained for arbitrary finite fields.

\subsection {The function field sieve}

The function field sieve is of particular interest in our case since it is related to the algorithm for curves of Section~\ref {ssec:l13curves}. Its starting point is the observation that the smoothness results of Section~\ref {ssec:framework} may be generalised by varying the size of the elements to be tested for smoothness and the smoothness bound. The following theorem is proved in \cite{eg07} for algebraic curves whose genus grow sufficiently fast compared to a power of $\log q$, but it holds in more generality.

\begin {theorem}
\label {th:smoothness}
Given an arithmetical formation of order $N$ as in Section~\ref {ssec:framework} in which smoothness is governed by the Dickmann--de Bruijn function, let $0 < \beta < \alpha \leqslant 1$ and $c$, $d > 0$. The probability that an element of size at most $\log L_N (\alpha, c)$ is smooth with respect to the factor base containing the $L_N (\beta, d)$ smallest primes is given by
\[
1 / L_N \left( \alpha - \beta, (\alpha - \beta) \frac {c}{d} + o (1) \right).
\]
\end {theorem}

The special case $\alpha = c = 1$ and $\beta = 1/2$ has been used in the previous section to prove complexities in $L (1/2)$. To reach a complexity of $L (1/3)$, this theorem opens only one direction: Since the factor base has to be written down, one may not exceed $\beta = 1/3$, whence the size of the elements to be decomposed has to be lowered to $\log L (2/3)$.

The function field sieve succeeds in this goal by representing the finite field $\F_{2^m}$, say, in two different ways: First of all, as before, by $\F_2 [X] / (f)$ with $f$ irreducible of degree $m$. Second, as residue class field of a place in a function field over $\F_2$, given by a $\Cc_{a, b}$ curve $\Cc : Y^a - F (X, Y) = 0$ with $b \approx a$. Suppose that the ideal $(f)$ is totally split in $\F_2 (\Cc)$, and let $\f = (f (X), Y - t (X))$ be an ideal of $\F_2 [\Cc]$ above $(f)$. Then the two homomorphisms with domain $\F_2 [X, Y]$, given on the one hand by the reduction $\psi : \F_2 [X, Y] \to \F_2 [\Cc]$ modulo the curve equation and on the other hand by the evaluation map $\varphi : \F_2 [X, Y] \to \F_2 [X]$, $Y \mapsto t (X)$, are compatible with the reductions modulo $\f$ and $f$:
\[
\begin {diagram}
\node [2]{\F_2 [X,Y]} \arrow {sw,l}{\psi}
\arrow{se,l}{\varphi : Y \mapsto t (X)} \\
\node {\F_2 [\Cc] = \rquo {\F_2 [X, Y]}{ (Y^a - F(X,Y))}} \arrow {s}
\node [2]{\F_2 [X]} \arrow {s} \\
\node {\rquo {\F_2 [\Cc]}{\f}}
\arrow [2]{e,t,..,-}{\simeq}
\node [2]{\rquo {\F_2 [X]}{(f)}}
\end {diagram}
\]

By drawing a random polynomial $w \in \F_2 [X, Y]$, one thus obtains a relation in $\F_{2^m}$ whenever both images under $\psi$ and $\varphi$ are smooth. Some technical complications stem from the fact that $\F_2 [\Cc]$ is in general not a principal domain, so that instead of decomposing $\psi (w)$, one is limited to decomposing the principal ideal it generates. Apart from this, decomposition on the function field side amounts to factoring the norm of $\psi (w)$ and is thus reduced again to factoring univariate polynomials.

The degree $a$ of the curve yields an additional degree of freedom; by choosing carefully the parameters, the degrees of the norm of $\psi (w)$ as well as of $\varphi (w)$ may be bounded by $\log L_{2^m} (2/3)$. At first sight, the situation is unfavourable since two smoothness conditions have to be satisfied simultaneously instead of only one. But by (\ref {eq:subexp}), this influences only the constant, while the lower degree of the polynomials to be factored acts on the more important first parameter of the subexponential function. The worse constant, however, implies that the algorithms in $L (1/3)$ are not immediately faster than those in $L (1/2)$, but only from a certain input size on. The complexity of $L (1/3)$ is only heuristic since it relies on the assumption that the norm of a polynomial in $\F_2 [\Cc]$ has the same smoothness probability as a random univariate polynomial of the same degree.

\subsection {An algorithm for a special class of curves}
\label {ssec:l13curves}

It is now a natural question to ask whether it is possible to reach a complexity of $L (1/3)$ also for the discrete logarithm problem in Jacobians of curves over finite fields. Given the analogies between finite fields and Jacobians used in Section~\ref {ssec:framework} to develop a unified theory for algorithms of complexity in $L (1/2)$, one might nourish some hope to similarly generalise the algorithms in $L (1/3)$ to curves. But the second way of representing $\F_{2^m}$ as a residue field in a function field (respectively, $\F_p$ as a residue field in a number field) does not seem to be parallelled in Jacobians. It is apparently impossible, for instance, to stack a second curve on top of the first one.

The solution to this problem presented in \cite{eg07} turns this apparent obstacle into an advantage: Indeed, it is suggested to work \textit {directly} with the curves that appear in the function field sieve. The algorithm is not limited to $\Cc_{a, b}$ curves. Let $a_0$ and $b_0$ be arbitrary positive constants. Consider a family of absolutely irreducible curves of genus $g$ over a finite field $\F_q$ of the form
\[
\Cc : Y^a + F (X, Y)
\]
with $F (X, Y) \in \F_q [X, Y]$ of degree $b$ in $X$ and at most $a-1$ in $Y$, where $a$ and $b$ are bounded by
\begin {equation}
\label {eq:ab}
a < a_0 g^{1/3} \Mcal^{-1/3} \text { and }
b < b_0 g^{2/3} \Mcal^{1/3}
\end {equation}
with $\Mcal = \frac {\log (g \log q)}{\log q} = \log_q (g \log q)$. To apply the smoothness result of Theorem~\ref {th:smoothness}, one furthermore has to impose that $g \geqslant (\log q)^\delta$ for some $\delta > 2$.

For instance, one may choose $a_0 > 0$ arbitrarily, fix $b_0 = \frac {2}{a_0}$ and consider $\Cc_{a, b}$ curves satisfying (\ref {eq:ab}); this ensures that we are not speaking about the empty set.

Relations are created in the same way as in Adleman--DeMarrais--Huang's algorithm of Section~\ref {ssec:l12hyperelliptic}: As principal divisors are zero in the Jacobian, it suffices to draw random polynomials $w = r (X) + s (X) Y$ and to verify whether their divisors are smooth; again, this amounts to factoring the norm of $w$ in $\F_q [X]$.

Choosing as factor base the $L_{q^g} (1/3, d)$ smallest prime divisors and the degrees of $r$ and $s$ as $c g^{1/3} \Mcal^{2/3}$, the following two properties hold:
\begin {itemize}
\item
First of all, the smoothness probability of the norm is (heuristically) given by
$1 / L (1/3, e/d + o (1))$ with $e = (a_0 c + b_0) / 3$.
\item
Second, the \textit {sieving space}, that is, the set from which the tuple $(r, s)$ is drawn, is sufficiently large. In fact, it would be possible to increase the smoothness probability by selecting $r$ and $s$ of even smaller (in the extreme case, constant) degrees; but then the number of choices for $w$ would be so restricted that one would not even obtain a single relation on average. As in other subexponential algorithms, one has to ensure that the number of random choices at one's disposition is at least as large as the number of smoothness tests carried out. This is the main obstacle to decreasing the complexity below $L (1/3)$.
\end {itemize}
By computing the Smith normal form of the relation matrix, one obtains the order and the structure of the Jacobian as a product of cyclic groups. With the optimal choice of the free parameters $c$ and $d$, the complexity becomes
\[
L_{q^g} \left( 1/3, \frac {4}{3} \sqrt {a_0 c + b_0} + o (1) \right),
\]
where $c$ is the positive solution of the quadratic equation
$c^2 - \frac {4}{9} a_0 c - \frac {4}{9} b_0 = 0$.

It remains to be seen how to compute discrete logarithms. One needs (as in the third stage of Algorithm~\ref {alg:l12}) an additional relation containing the divisor $Q$ whose logarithm is sought. Unfortunately, the size of $Q$ cannot be controlled, and chances are that it is of order $\log L (1)$ rather than $\log L (2/3)$. Perturbing $Q$ randomly by elements of the factor base, the smoothness theorem~\ref {th:smoothness} implies that in average time $L (1/3)$, one may obtain a relation containing $Q$ and prime divisors $Q_i$ of size $\log L (2/3)$. It is possible to use an approach called \textit {special $q$ descent} in the context of factorisation, creating for each $Q_i$  a relation containing it by considering functions $w = r (X) + s (X) Y$ passing through $Q_i$. But in order to have a reasonable chance of finding a relation, one needs to arrange some freedom for the degrees of $r$ and $s$; with the additional restriction of passing through one of the $Q_i$, one again has to decompose a divisor of degree $\log L (1)$, and the process turns in circles.

The solution suggested in \cite{eg07} consists in relaxing slightly the constraint on the running time. Let thus $\varepsilon > 0$ be fixed. In time $L (1/3 + \varepsilon)$, one may create a relation containing $Q$ and further prime divisors $Q_i$ of degree $\log L (2/3 - \varepsilon)$. For each $Q_i$, a special $q$ descent allows to replace it in time $L (1/3 + \varepsilon)$ by a linear combination of prime divisors $Q_{i, j}$ of degree $\log L (2/3 - 2 \varepsilon)$; these $Q_{i, j}$ are again treated by a special $q$ descent, and so forth. Whenever the degree of a $Q_{i, j, \ldots}$ drops below the barrier of $L (1/3 + \varepsilon)$, the descent returns primes of degree $\log L (1/3)$, which are elements of the factor base, and the process terminates.

This descent approach creates a tree in which all nodes have a degree in $O (g)$, whose height is bounded by $1 / (3 \varepsilon)$, and whose leaves are in the factor base. As $\varepsilon$ is a constant, the number of nodes in the tree is polynomial in $g$ and thus generously covered by any subexponential function. So the following result holds:

\begin {theorem}[heuristic]
Let there be given a family of curves $\Cc$ as above, satisfying in particular (\ref {eq:ab}) and $g \geqslant (\log q)^\delta$ for some $\delta > 2$, and let $\varepsilon > 0$. Assuming heuristically that the divisors encountered during the algorithm have the same smoothness probability of Theorem~\ref {th:smoothness} as random divisors of the same degree, discrete logarithms in the Jacobian of $\Cc$ can be computed in time $L_{q^g} (1/3 + \varepsilon, o (1))$.
\end {theorem}

Concerning the constant of the subexponential complexity, it suffices to note that the existence of an algorithm in $L (1/3 + \varepsilon / 2, c)$ for some constant $c$  allows to reach $L (1/3 + \varepsilon, o (1))$ by (\ref {eq:subexp}).

The degrees $a$ in $X$ and $b$ in $Y$ of the curve may be balanced differently. Letting $a \approx g^\alpha$ and $b \approx g^{1 - \alpha}$ for some $\alpha$ between $1/3$ and $1/2$, the algorithm for computing the group structure remains of complexity $L (1/3)$ (with a different constant depending on $\alpha$), while the time for computing discrete logarithms becomes $L (\alpha + \varepsilon)$. When $\alpha$ drops below $1/3$, also the group structure computation becomes slower than $L (1/3)$; its complexity turns out to be $L (x (\alpha))$ for $x (\alpha) \in [1/3, 1/2]$ and, in particular, $L (1/2)$ for hyperelliptic curves. This is apparently the first natural occurrence of an algorithm with a subexponential complexity in which the first parameter is different from $1/3$ and $1/2$.

\subsection {The low genus case}
\label {ssec:l13low}

In the spirit of Section~\ref {ssec:l12low}, Diem has considered in \cite{die06} a particular class of low genus curves, in which discrete logarithms are easier to compute. His ideas are independent of the algorithm of complexity $L (1/3)$ presented in the previous section, but it turns out that the gain with respect to general curves is in both cases due to a curve degree that is comparatively small for a given genus. Let again be given a family of curves with $q$ tending to infinity, this time represented by plane models of fixed degree $d$ instead of a fixed genus. The factor base is formed, as in Section~\ref {ssec:l12low}, by the prime divisors of degree~$1$, otherwise said, the rational points on the curve. Relations are created, as in the algorithm of Section~\ref {ssec:l13curves} and as in Adleman--DeMarrais--Huang's algorithm, by computing divisors of polynomials; so like these two algorithms, Diem's approach is heuristic. He considers furthermore only the simplest polynomials, namely lines. By Bézout's theorem, they intersect the curve in $d$ points (that may have coordinates in an extension field); so a relation is obtained whenever a polynomial of degree~$d$ factors into linear factors over the base field, as opposed to Section~\ref {ssec:l12low}, in which the polynomial was of degree~$g$. Using the double prime variation, one obtains an algorithm of heuristic complexity $O (q^{2 - 2 / d})$, measured in arithmetic operations. Diem suggests an additional trick to lower the complexity; he restricts to lines drawn between two points that are already in the factor base. Then a polynomial of degree only $d-2$ has to split into linear factors to yield a relation, and the complexity of the discrete logarithm algorithm drops to $O (q^{2 - 2 / (d-2)})$. This algorithm is preferable to the one described in Section~\ref {ssec:l12low} whenever $d - 2 < g$. Hyperelliptic curves are not concerned, but the impact on $\Cc_{a,b}$ curves with $3 \leqslant a < b$ is dramatic. The equations $d = b$ and $g = \frac {(a - 1)(b - 1)}{2}$ imply that discrete logarithms are obtained with $O (q^{2 - 2 (a - 1) / (2 g - (a - 1))})$ operations. In particular, in the case $a = 3$ and $b = 4$ the complexity is $O (q)$; so the discrete logarithm problem in non-hyperelliptic $\Cc_{a,b}$ curves of genus~$3$ is not harder than in hyperelliptic curves of genus~$2$ \textit {defined over the same finite field}, while the bit length of the group order is $50~\%$ higher and the arithmetic is considerably more involved. This result implies that non-hyperelliptic curves are not suited for the implementation of discrete logarithm based cryptosystems.

\section {Implementations}

The latest data points for computing discrete logarithms with a generic algorithm are from 2002 and 2004 and concern elliptic curves over prime fields and fields of characteristic~$2$ of $109$~bits \cite{cer02,cer04ann}; the 2004 computation involved $2600$~processors running over $17$~months.

A subexponential algorithm for hyperelliptic curves has first been implemented by Flassenberg and Paulus \cite{fp99sie}. Their largest example, a curve of genus~$12$ over $\F_{11}$, is far from reaching a cryptographic parameter size; since the cardinalities of these high genus curves were unknown, the authors had to resort to expensive Hermite normal form computation instead of solving a sparse linear system. Gaudry reports on an implementation of the algorithm of Section~\ref {ssec:l12low} (without large primes) in \cite{gau00alg}; his largest examples, curves of genus~$6$ over $\F_{5026243}$ respectively $\F_{2^{23}}$, surpass the generic record of the previous paragraph and are very close to cryptographic group orders.

The algorithm of Section~\ref{ssec:l13low}, including the double large prime variation, has been implemented by Diem and Thomé for a $\Cc_{3,4}$ curve of genus~$3$ over $\F_{2^{31}}$, see \cite{dt07}. Their computation taking only a few days with the relation collection carried out on a single CPU, this should rather be seen as a proof of concept for the algorithm than as a benchmark on what is achievable today. The authors estimate that discrete logarithms on a $\Cc_{3,4}$ curve with a group order of $111$~bits could be obtained by an effort comparable to that of factoring a $664$~bit RSA integer.

\section {Future research}

The algorithm of Section~\ref {ssec:l13curves} of complexity $L (1/3 + \varepsilon)$ for computing discrete logarithms in certain curves opens a new direction of research. During the 10th Workshop on Elliptic Curve Cryptography (ECC 2006), Diem has announced an algorithm of complexity $L (1/3)$ inspired by these ideas, but with a quite different point of view \cite{die06ecc}; for the time being, it is unclear whether his class of curves is different from the one considered in Section~\ref {ssec:l13curves}. It would be interesting to obtain a complete classification of the curves that are subject to a subexponential attack of complexity better than $L (1/2)$.

In a recent preprint \cite{smi07}, Smith has found a novel attack on certain hyperelliptic curves of genus~$3$. He explicitly computes an isogeny to a non-hyperelliptic curve of genus~$3$, which allows to transport the discrete logarithm problem and to solve it via the algorithm of Section~\ref {ssec:l13low}. Heuristically, the attack applies to about one out of five hyperelliptic curves of genus~$3$. However, by considering more general isogenies, it appears likely that the result could be extended to other curves, which would cast further doubt on the use of genus~$3$ curves in cryptography.

\bibliographystyle {amsplain}
\providecommand{\bysame}{\leavevmode\hbox to3em{\hrulefill}\thinspace}
\providecommand{\MR}{\relax\ifhmode\unskip\space\fi MR }
\providecommand{\MRhref}[2]{%
  \href{http://www.ams.org/mathscinet-getitem?mr=#1}{#2}
}
\providecommand{\href}[2]{#2}

\end {document}

%% file: function.pstex_t
\begin{picture}(0,0)%
\includegraphics{function.pstex}%
\end{picture}%
\setlength{\unitlength}{3552sp}%
\begingroup\makeatletter\ifx\SetFigFontNFSS\undefined%
\gdef\SetFigFontNFSS#1#2#3#4#5{%
  \reset@font\fontsize{#1}{#2pt}%
  \fontfamily{#3}\fontseries{#4}\fontshape{#5}%
  \selectfont}%
\fi\endgroup%
\begin{picture}(5839,3579)(1317,-4012)
\put(1563,-3648){\makebox(0,0)[rb]{\smash{{\SetFigFontNFSS{9}{10.8}{\familydefault}{\mddefault}{\updefault}0}}}}
\put(1563,-3340){\makebox(0,0)[rb]{\smash{{\SetFigFontNFSS{9}{10.8}{\familydefault}{\mddefault}{\updefault}20}}}}
\put(1563,-3033){\makebox(0,0)[rb]{\smash{{\SetFigFontNFSS{9}{10.8}{\familydefault}{\mddefault}{\updefault}40}}}}
\put(1563,-2725){\makebox(0,0)[rb]{\smash{{\SetFigFontNFSS{9}{10.8}{\familydefault}{\mddefault}{\updefault}60}}}}
\put(1563,-2418){\makebox(0,0)[rb]{\smash{{\SetFigFontNFSS{9}{10.8}{\familydefault}{\mddefault}{\updefault}80}}}}
\put(1563,-2110){\makebox(0,0)[rb]{\smash{{\SetFigFontNFSS{9}{10.8}{\familydefault}{\mddefault}{\updefault}100}}}}
\put(1563,-1803){\makebox(0,0)[rb]{\smash{{\SetFigFontNFSS{9}{10.8}{\familydefault}{\mddefault}{\updefault}120}}}}
\put(1563,-1495){\makebox(0,0)[rb]{\smash{{\SetFigFontNFSS{9}{10.8}{\familydefault}{\mddefault}{\updefault}140}}}}
\put(1563,-1188){\makebox(0,0)[rb]{\smash{{\SetFigFontNFSS{9}{10.8}{\familydefault}{\mddefault}{\updefault}160}}}}
\put(1563,-880){\makebox(0,0)[rb]{\smash{{\SetFigFontNFSS{9}{10.8}{\familydefault}{\mddefault}{\updefault}180}}}}
\put(1563,-573){\makebox(0,0)[rb]{\smash{{\SetFigFontNFSS{9}{10.8}{\familydefault}{\mddefault}{\updefault}200}}}}
\put(2149,-3773){\makebox(0,0)[b]{\smash{{\SetFigFontNFSS{9}{10.8}{\familydefault}{\mddefault}{\updefault}200}}}}
\put(2686,-3773){\makebox(0,0)[b]{\smash{{\SetFigFontNFSS{9}{10.8}{\familydefault}{\mddefault}{\updefault}400}}}}
\put(3224,-3773){\makebox(0,0)[b]{\smash{{\SetFigFontNFSS{9}{10.8}{\familydefault}{\mddefault}{\updefault}600}}}}
\put(3762,-3773){\makebox(0,0)[b]{\smash{{\SetFigFontNFSS{9}{10.8}{\familydefault}{\mddefault}{\updefault}800}}}}
\put(4300,-3773){\makebox(0,0)[b]{\smash{{\SetFigFontNFSS{9}{10.8}{\familydefault}{\mddefault}{\updefault}1000}}}}
\put(4837,-3773){\makebox(0,0)[b]{\smash{{\SetFigFontNFSS{9}{10.8}{\familydefault}{\mddefault}{\updefault}1200}}}}
\put(5375,-3773){\makebox(0,0)[b]{\smash{{\SetFigFontNFSS{9}{10.8}{\familydefault}{\mddefault}{\updefault}1400}}}}
\put(5913,-3773){\makebox(0,0)[b]{\smash{{\SetFigFontNFSS{9}{10.8}{\familydefault}{\mddefault}{\updefault}1600}}}}
\put(6450,-3773){\makebox(0,0)[b]{\smash{{\SetFigFontNFSS{9}{10.8}{\familydefault}{\mddefault}{\updefault}1800}}}}
\put(6988,-3773){\makebox(0,0)[b]{\smash{{\SetFigFontNFSS{9}{10.8}{\familydefault}{\mddefault}{\updefault}2000}}}}
\put(4313,-3960){\makebox(0,0)[b]{\smash{{\SetFigFontNFSS{9}{10.8}{\familydefault}{\mddefault}{\updefault}$\log_2 N$}}}}
\put(4963,-2988){\makebox(0,0)[lb]{\smash{{\SetFigFontNFSS{9}{10.8}{\familydefault}{\mddefault}{\updefault}$\log_2 \sqrt N$}}}}
\put(4963,-3222){\makebox(0,0)[lb]{\smash{{\SetFigFontNFSS{9}{10.8}{\familydefault}{\mddefault}{\updefault}$\log_2 L_N(1/2,\,\sqrt2)$}}}}
\put(4963,-3456){\makebox(0,0)[lb]{\smash{{\SetFigFontNFSS{9}{10.8}{\familydefault}{\mddefault}{\updefault}$\log_2 L_N(1/3,\,2)$}}}}
\end{picture}%